\documentstyle[11pt,paspconf,epsf]{article}

\begin{document}

\title{An ISO Survey of PAH Features in Ultraluminous Infrared Galaxies}
\author{D. Lutz, R. Genzel, D. Rigopoulou, H.W.W.Spoon, D.Tran}
\affil{MPE, Postfach 1603, 85740 Garching, Germany}
\author{A.F.M. Moorwood}
\affil{ESO, Karl-Schwarzschildstra\ss\/e 1, 85748 Garching, Germany}

\begin{abstract}
We have obtained ISOPHOT-S low resolution mid-infrared spectra of a sample
of 60 Ultraluminous Infrared Galaxies (ULIRGs). We use the strength of the
`PAH' mid-infrared features as a discriminator between starburst and AGN
activity, and to probe for evolutionary effects. Observed ratios of PAH 
features in ULIRGs differ slightly from
those in lower luminosity starbursts. We suggest that such PAH ratio changes 
relate to the conditions in the interstellar medium in these galaxies, and
in particular to extinction. 
\end{abstract}


\section{Introduction}

The nature of ultraluminous infrared galaxies has been the
subject of lively debate since their discovery by IRAS more than a
decade ago. Although evidence for both starburst and AGN activity in
ULIRGs has accumulated during this period, the question as to which
generally dominates the luminosity has remained largely unsolved,
mainly due to observational difficulties associated with their
large dust obscuration.

\begin{figure}
\plotfiddle{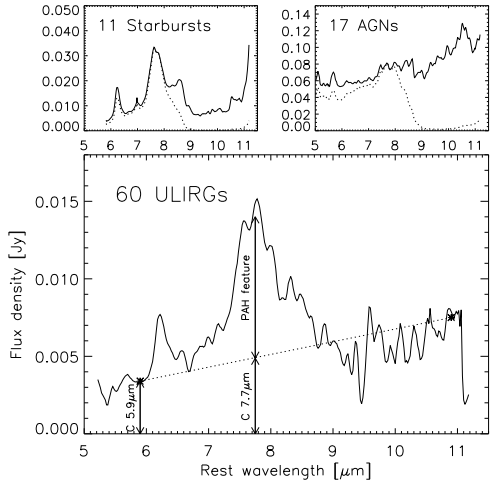}{6.5cm}{0.}{40.}{40.}{-100.}{0.}
\caption{Average ISOPHOT-S spectrum of all ULIRGs observed,
individually scaled to $S_{60}$=1Jy. Average 
spectra of starburst galaxies and AGNs are added for comparison. The dashed lines
represent these spectra after applying an additional A$_V$=50 foreground
extinction}
\end{figure}

With the advent of ISO, sensitive mid-infrared spectroscopy became
available as a new tool capable of penetrating the obscuring
dust. Fine structure line and PAH feature observations with SWS and
ISOPHOT-S of a sample of 15 bright ULIRGs suggest that most are
starburst-powered (Genzel et al. 1998). However, this sample is too small
to search for luminosity or evolutionary effects. Using only PHT-S it
has subsequently proved possible to extend to fainter sources and
increase to 60 the number of ULIRGs observed over the wavelength range
containing the 6.2, 7.7, 8.6, and 11.3$\mu$m features commonly
attributed to polycyclic aromatic hydrocarbons (PAH). ISO data
confirm groundbased observations of these and a companion at 3.3$\mu$m 
which first demonstrated
that these features are strong in starburst galaxies but weak or
absent in classical AGNs. We have therefore
used the line to continuum ratio of the most prominent,
7.7$\mu$m, feature as our primary discriminator between starburst and
AGN activity.

\section{A PAH Survey of ULIRGs}

Our sample of 60 ULIRGs is drawn from the 1.2\,Jy survey
(Fisher et al. 1995).  No infrared color criteria were applied to avoid
biasing the sample in AGN content.  
The average of all 60 ULIRG spectra, individually scaled to
S$_{60}$=1Jy to give all sources equal weight
(Fig.~1), clearly shows the PAH features at 6.2, 7.7,
and 8.6$\mu$m but relatively weak continuum. Comparison with the
starburst and AGN templates provides a first and direct indication
that ULIRGs are, on average, starburst-like.  Fig.~1
also illustrates our method for extracting PAH and continuum data from
the individual spectra. We find that
\begin{itemize}
\item About 80\% of all the ULIRGs are found to be predominantly powered by star
formation but the fraction of AGN powered objects increases with
luminosity. Whereas  only about 15\% of ULIRGs at luminosities below $2\times
10^{12} L_{\sun}$ are AGN powered this fraction reaches
about half at higher luminosity.
\item The PAH feature-to-continuum ratio is anticorrelated with the ratio of
feature-free 5.9$\mu$m continuum to the IRAS 60$\mu$m continuum,
confirming suggestions that strong mid-IR continuum is a prime AGN
signature.  The location of starburst-dominated ULIRGs in such a
diagram is consistent with previous ISO-SWS spectroscopy which implies
significant extinction even in the mid-infrared. 
\item We have searched for indications that ULIRGs which are advanced
mergers might be more AGN-like, as postulated by the classical
evolutionary scenario. No such trend has been found amongst
those objects for which near infrared images are available to assess
their likely merger status.
\end{itemize}
See Lutz et al. (1998) for a discussion of these results.

\section{PAH ratios as indicators of extinction and ISM conditions}

\begin{figure}
\plotfiddle{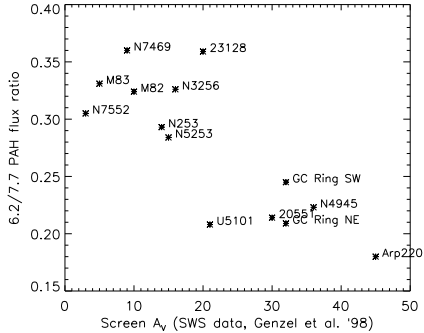}{4.6cm}{0.}{35.}{35.}{-90.}{0.}
\caption{Anticorrelation between the ratio of the 6.2 and 7.7$\mu$m PAH features
and extinction}
\end{figure}

\begin{figure}
\vspace{7.cm}
\caption{A scenario of how extinction and intense radiation fields may affect
mid-IR PAH spectra of galaxies. NGC 3256 represents starbursts and normal
galaxies.}
\end{figure}

Observed ratios of the PAH features in ULIRGs differ slightly from those
in lower luminosity starbursts which exhibit very homogeneous
PAH properties (Fig.~1). The ratio of the 6.2/7.7$\mu$m
features is lower and the 8.6$\mu$m feature is a shoulder to the 7.7$\mu$m
one rather than a well-defined feature. This behaviour is more pronounced
in some individual spectra, e.g. for Arp 220. 
One possibility is
that the weakness of the 6.2$\mu$m feature reflects the unusual
conditions of the ULIRG interstellar medium, as sometimes observed
in galactic sources. For
compact H\,II regions, Roelfsema et al. (1998)
find 6.2/7.7 ratios which on
average are lower than for `typical' H\,II regions, and for
some sources high 8.6/7.7 ratios which might be linked to an
intense radiation field. 

A second effect is
the influence of strong extinction which is already suggested by the
similarity of the average ULIRG spectrum and the obscured starburst spectrum
of Figure~1. Extinction suppresses the 6.2, 8.6,
and 11.3 features in comparison to the one at 7.7$\mu$m. Figure~2 demonstrates
that this effect is indeeed at work: There is an anticorrelation between
6.2/7.7 ratio and extinction estimated from independent ISO-SWS spectroscopy
(Genzel et al. 1998 values, converted to screen case).
Of the starburst templates shown,
extinction approaches ULIRG levels only for NGC\,4945 and the
molecular ring encircling the center of our Galaxy. Interestingly,
these are the only spectra in that group which also show low
6.2/7.7$\mu$m flux ratios. Further, as in our average ULIRG spectrum,
their 8.6$\mu$m features appear as a shoulder to the 7.7$\mu$m feature rather
than as a separate feature, due to suppression in the wings of the silicate 
absorption feature.

While extinction appears to be the prime cause for unusual PAH ratios
in the ULIRGs and source like NGC 4945, we suggest that both extinction and
intense radiation fields govern the appearance of the mid-infrared PAH spectra
of galaxies
(Fig. 3). Galaxies like the starbursting dwarf NGC 5253 and the circumnuclear
region of the Seyfert 2 NGC\,1068 show a broadish 7.7-8.6 PAH complex with
high 8.6/7.7 ratio, just like the inner parts of HII regions (Roelfsema et al.
1998, Verstraete et al. 1996).

\end{document}